\documentclass[runningheads]{llncs}

\usepackage[T1]{fontenc}
\def\doi#1{\href{https://doi.org/\detokenize{#1}}{\url{https://doi.org/\detokenize{#1}}}}
\usepackage{graphicx}
\usepackage{amssymb}%for huati R

\usepackage{listings}
\begin{document}
\title{An Automatic Cascaded Model for Hemorrhagic Stroke Segmentation and Hemorrhagic Volume Estimation}

\author{Weijin Xu\inst{1} \and
Zhuang Sha\inst{2} \and
Huihua Yang\inst{1} \and
Rongcai Jiang\inst{2} \and
Zhanying Li\inst{3} \and
Wentao Liu\inst{1} \and
Ruisheng Su\inst{4}
}
\institute{Beijing University of Posts and Telecommunications, China \and
Department of Neurosurgery, Tianjin Medical University General Hospital, China \and
Department of Neurosurgery, Kailuan General Hospital, China \and
Erasmus MC, Netherlands\\
\email{xwj1994@bupt.edu.cn}
\footnote{Weijin Xu and Zhuang Sha contribute equally, Huihua Yang and Rongcai Jiang are co corresponding authors.}\\
}

\maketitle              % typeset the header of the contribution
\begin{abstract}
Hemorrhagic Stroke (HS) has a rapid onset and is a serious condition that poses a great health threat. Promptly and accurately delineating the bleeding region and estimating the volume of bleeding in Computer Tomography (CT) images can assist clinicians in treatment planning, leading to improved treatment outcomes for patients. In this paper, a cascaded 3D model is constructed based on UNet to perform a two-stage segmentation of the hemorrhage area in CT images from rough to fine, and the hemorrhage volume is automatically calculated from the segmented area. On a dataset with 341 cases of hemorrhagic stroke CT scans, the proposed model provides high-quality segmentation outcome with higher accuracy (DSC 85.66\%) and better computation efficiency (6.2 second per sample) when compared to the traditional Tada formula with respect to hemorrhage volume estimation.

% \textbf{Background} Intracranial hemorrhage (ICH) is a serious health problem requiring prompt and intensive medical treatment.

% \textbf{Objective} To develop an automatic segmentation framework based on deep learning to assist neurologists in delineating the area of cerebral hemorrhage and evaluating the amount of bleeding in CT images.

% \textbf{Methods} The framework is based on a encoder-decoder manner for CT image segmentation, which is fully convolutional and can be optimized end-to-end. Moreover, a channel-attention module is employed to adjustment the importance of different feature representation. The proposed network was trained on training set with 238 CT and tested on a blind testset with 103 CT images.

% \textbf{Results}

% \textbf{Conclusions} This deep neural network assists in successfully segmenting cerebral hemorrhage area from CT images and evaluating the amount of bleeding in CT images, and can be used in clinical practice, which has the potential to be applied in medical practice.

\keywords{Hemorrhagic stroke  \and CT \and Hemorrhagic volume estimation \and Deep learning \and Medical image segmentation}
\end{abstract}

\section{Introduction}\label{introduction}

\begin{figure}[hbtp]
\centering
\includegraphics[scale=0.275]{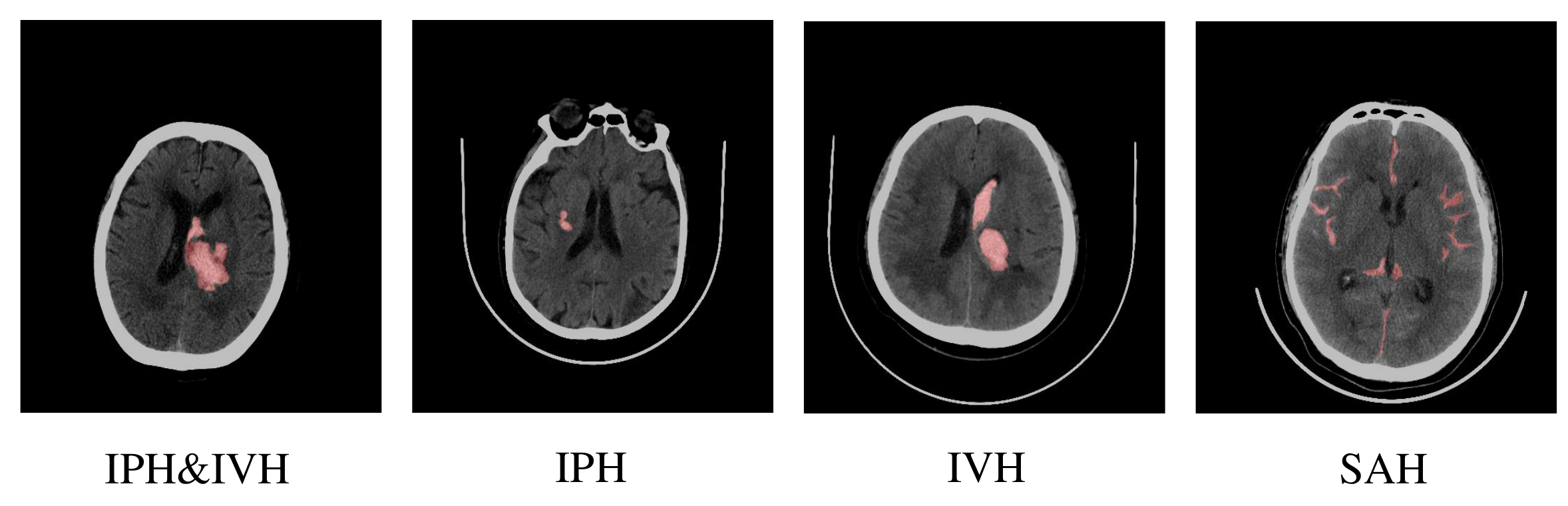}
\caption{Sample displays of different types of bleeding, with red masks indicating the bleeding area, where IPH \& IVH means the hybrid hemorrhage type of IPH and IVH. Best view in color.} \label{fig:HS_show}
\end{figure}

Stroke is the world's number one deadly and second most disabling disease, posing a huge threat to people's health and a huge burden to the healthcare system~\cite{stroke}. Stroke can be divided into Ischemic Stroke (IS) and Hemorrhagic Stroke (HS), with a high incidence (about 70\%) of IS and a relatively low incidence (about 30\%) of HS. However, HS is characterized by its rapid onset, grave nature, and significantly higher mortality rate. HS refers to cerebral hemorrhage caused by non-traumatic rupture of blood vessels in the brain parenchyma, and it contains three types of brain hemorrhage: subarachnoid hemorrhage (SAH), intraparenchymal hemorrhage (IPH), and intraventricular hemorrhage (IVH), as shown in Fig~\ref{fig:HS_show}.
The most common causes are hypertension, cerebral atherosclerosis, intracranial vascular malformations, which are often triggered by exertion and emotional excitement~\cite{donkor2018stroke}. The prime treatment time for the onset of hemorrhagic stroke is within 3 hours after the onset. Therefore, timely and rapid determination of the amount, site, and contour of bleeding can bring better treatment results, fewer complications, and less severe sequelae to the patient~\cite{Timely}.

The choice of diagnostic methods in each specific case strongly depends not only on their applicability (availability, contraindications, patient’s condition, etc.), but also on the time of symptom onset. Any delay in medical care increases the risk of severe consequences and death~\cite{HMOE-Net}. Computer Tomography (CT) is used as a first-line diagnostic modality because of its short imaging time, ease of acquisition, high sensitivity to bleeding, high resolution, and clear anatomic relationships. The physicians need to inspect the CT scan and make the appropriate treatment plan depending on the type of bleeding, the area of bleeding, and the size of the bleeding, which takes effort and may present human error in identifying hemorrhages. Therefore, this paper constructs an automated pipeline based on convolutional neural networks (CNNs) to outline the bleeding area in CT, which may reduce the workload of physicians, speed up the diagnosis process, and bring better treatment for patients.

% Intracranial hemorrhage (ICH) has a very high mortality rate and requires prompt diagnosis and further treatment, as tissue changes in the ischemic penumbra may be reversible, and requires urgent neurosurgical evacuation. 

% Add some statistics of the effect of ICH with people.

% The choice of diagnostic methods in each specific case strongly depends not only on its applicability (availability, contraindications, patient’s condition, etc.), but also on the time of symptoms onset. Any delay in medical care increases the risk of severe consequences and death. Computer Tomography (CT) is used as a first-line diagnostic modality because of its short imaging time, ease of acquisition, high sensitivity to bleeding, high resolution, and clear anatomic relationships. ICH has five types: Intraventricular Hemorrhage (IVH), Intraparenchymal Hemorrhage (IPH), Subarachnoid Hemorrhage (SAH), Epidural Hemorrhage (EDH), and Subdural Hemorrhage (SDH).  The doctor needs to inspect the CT scan and make the appropriate treatment plan depending on the type of bleeding, the area of bleeding, and the size of the bleeding, which takes an effort and may present human error in identifying hemorrhages. Therefore, this paper constructs an automated pipeline based on convolutional neural network to outline the bleeding area and predict the bleeding category in CT, which can reduce the workload of doctors, speed up the diagnosis process, and bring better treatment for patients.

\section{Related Work}\label{RW}
In~\cite{lee2020detection}, a novel artifcial neural network framework method is proposed for detecting the presence or absence of intracranial hemorrhage (ICH) and classifying the type of hemorrhage on CT images of the brain, achieving a performance of AUC 0.859 for detection and AUC 0.903 for classification among 250 CT images (100 healthy, 150 hemorrhage), respectively. RADnet~\cite{RADnet} utilizes a 2D CNN to extract the information from the frame-by-frame CT images, then uses Bil-LSTM to fuse the inter-frame dimension-only information, and finally determines whether there is a cerebral hemorrhage in the input CT sequence. In~\cite{RSNA}, a publicly available 2D dataset of CT frames (752,803 dcms), RSNA, for hemorrhage type classification, RSNA, was proposed, and a baseline ResNet-based classification method was constructed, which ultimately achieved a 93\% multi-category classification accuracy. IHA-Net~\cite{IHA-Net} proposed a residual hybrid atrous module to capture features for multiple receptive fields of different sizes and utilizes the idea of deep supervision to add constraints to the middle layer of the network, which makes the network converge faster. HMOE-Net~\cite{HMOE-Net} proposed a shallow-deep feature extraction network to deal with hybrid multi-scale object features. Although these methods have achieved good performance, none of them have been compared with clinically used methods for obtaining cerebral hemorrhage volumes. In this paper, the results of segmentation are converted to hemorrhage volume and compared with clinically used methods, demonstrating the great advantages of this method.

\section{Dataset}\label{Dataset}
The dataset used in this paper is composed of 341 CT samples that were retrospectively collected from 341 patients of two cohorts that were imaged by GE and Philips CT scan devices between June 2021 and March 2022. All patient-related information was erased. One physician with 5 years of experience delineated the bleeding area slice-by-slice, and another chief physician with 15 years of experience further revised and finalized the data. The background pixels are assigned a value of 0, and the bleeding area is assigned a value of 1. Moreover, the standard windows of 90 Hounsfield units (HU), and center level of 40 HU are employed on all CT samples to limit visual variation. Randomly, the 70\%(238) of the 341 CT samples are divided into training sets and 30\%(103) into test sets. The training set is used to train the pipeline, and the performance in the test set is used for performance comparation. 
\begin{figure}[hbtp]
\centering
\includegraphics[scale=0.375]{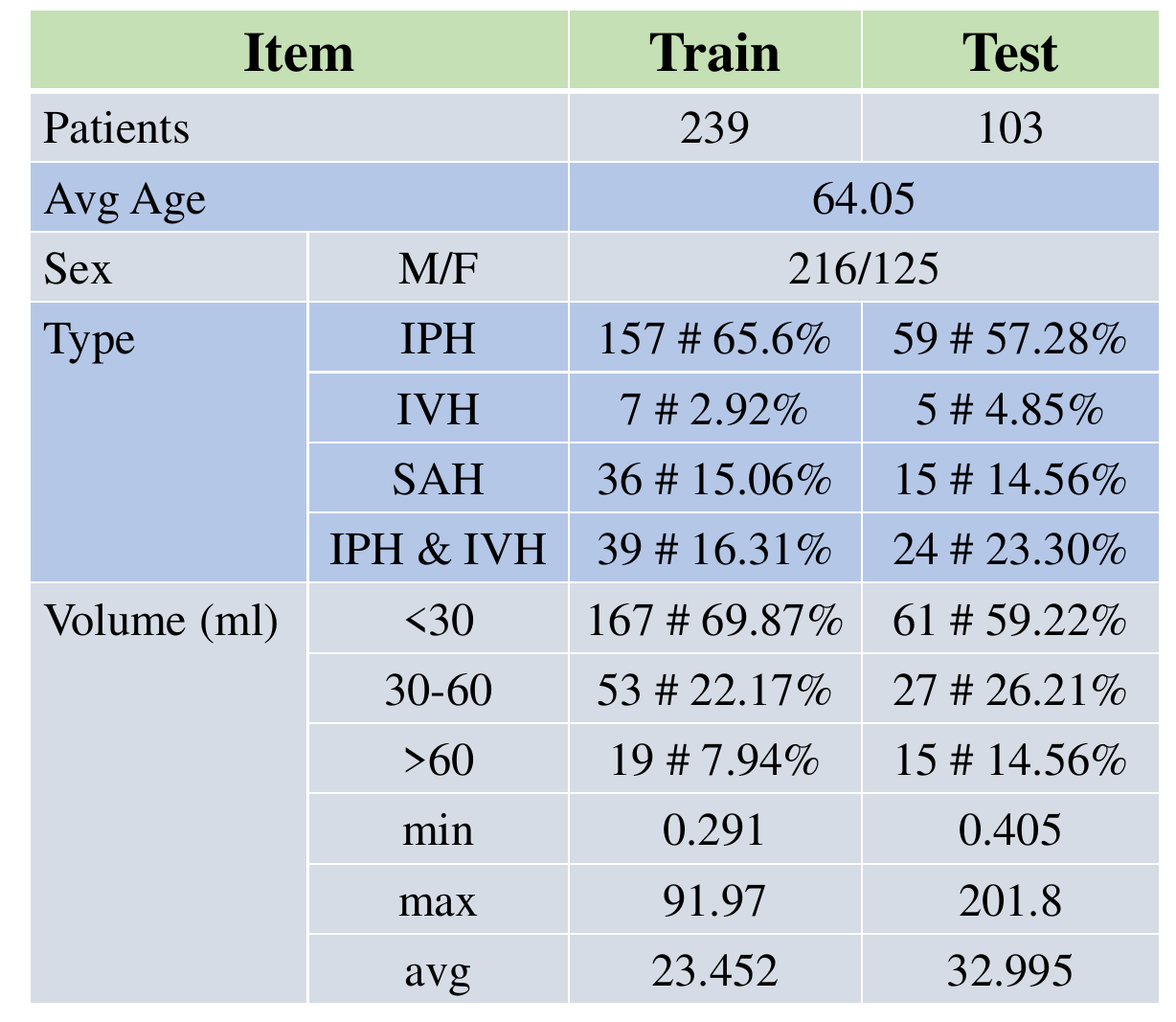}
\caption{Dataset characteristics. Best view in color.} \label{fig:static}
\end{figure}
The slice number varies from 20 to 47, with the average slice number of 28.13. The height varies from 508 to 512, and the width varies from 508 to 586. The median slice image size is 28$\times$483$\times$483, with the median spacing as 5$\times$0.518$\times$0.518 $mm^{3}$. The dataset characteristics are shown in Fig~\ref{fig:static}.

\section{Method}\label{Method}
The proposed pipeline is shown in Fig.~\ref{fig:Net}, it is constructed by Pytorch-2.0.0, which takes the 3D CT scans as input and outputs the 3D mask with the same size as input. The pipeline has two key components:1) the cascaded 3D encoder-decoder fully convolutional model to segment hemorrhages from coarse to fine; and 2) the deep supervision mechanism to constrain the training in the intermediate process to speed up model convergence.

% \subsection{Data augmentation}
% Convolutional neural networks are data-driven and require large amounts of data to extract semantic information and construct feature representations, so we employed a variety of data augmentation strategies, including: 1) random rotation between -30 degrees and 30 degrees; 2) random vertical flipping and horizontal flipping; 3) randomly adding Gaussian noise with 50\% probability; 4) random Gaussian smoothing with 50\% probability; 5) random contrast adjustment with 50\% probability; 6) randomly cropping out a region with a shape of 16$\times$320$\times$320, where 16 represents the slice depth, and 320$\times$320 indicates the visual size; 7) Normalize the input scan to the 0–1 range by subtracting the mean value and dividing the standard deviation value.

\subsection{Cascaded 3D model}
\begin{figure}[h]
\centering
\includegraphics[scale=0.3]{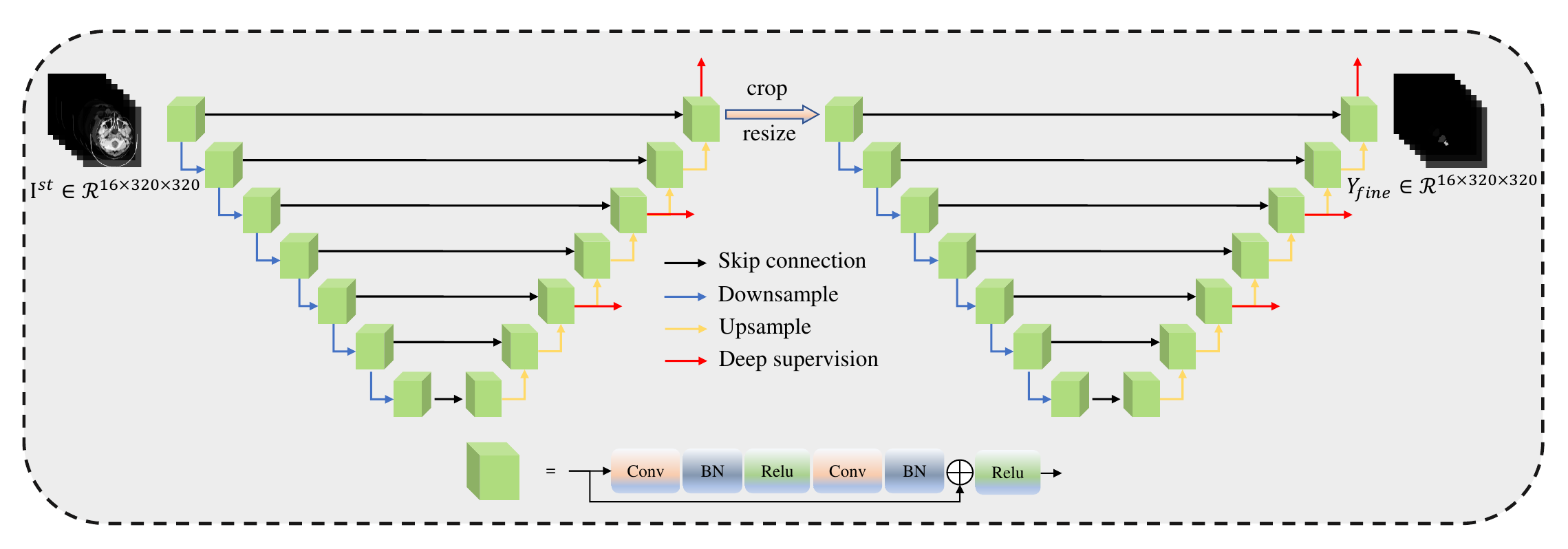}
    \caption{The diagram of Cascaded Encoder-Decoder convolutional model} \label{fig:Net}
\end{figure}

The cascaded 3D model is based on the popular encoder-decoder structure, UNet~\cite{UNet}, as shown in Fig.~\ref{fig:Net}. It has two stages, the first stage segments the coarse bleeding area, and this area is enlarged, cropped, resized, and fed to the second stage to get the precise hemorrhage masks. 
The structure of the first stage and the second stage is the same, but weights are not shared. The encoder and decoder both have six levels as shown in Fig.~\ref{fig:Net}, each level has two conv layers, each conv layer is followed by a batch-normalization layer~\cite{BN} to control gradient explosion and prevent gradient disappearance, and a Relu layer~\cite{relu} to add more nonlinearity. Moreover, the residual connection~\cite{ResUNet} is also employed to accelerate gradient propagation. At the same level of the encoder and decoder, the skip connection is used to transfer the low-level texture features lost in the downsampling process to the decoder, which can help the decoder reconstruct high-level semantic features. In the first stage, the cropped CT patch $I_{st} \in \mathcal{R}^{16\times320\times320}$ is first sent to the encoder to be downsampled gradually and extract feautres. However, the inter-slice dimension is only downsampled 2 times, and the intra-slice dimension is downsampled 6 times, which generates the bottleneck feature $F_b \in \mathcal{R}^{4\times5\times5}$. Then, $F_b$ is sent to the decoder to upsample and reconstruct the coarse prediction $Y_{coarse} \in \mathcal{R}^{16\times320\times320}$. In the second stage, the coarse bleeding area in $Y_{coarse}$ is cropped and resized as $I_{nd} \in \mathcal{R}^{16\times320\times320}$, which is processed as $I_{st}$ to get the fine prediction $Y_{fine} \in \mathcal{R}^{16\times320\times320}$. $Y_{fine}$ is compared with the label $Y_{gt}$ to calculate the metrics.

\subsection{Deep supervision mechanism}
Deep supervision~\cite{deepsuper} has been proven to be a plug-and-play technique in training convolutional models, it can regularize the feature extraction and reconstruction of the model and speed up the convergence. Therefore, as the red arrow shown in Fig.~\ref{fig:Net}, we also calculate losses during the upsampling process in the middle, besides the losses in the final predictions in each stage. Then, all these losses are added together to get the final loss. In this work, Dice coefficient (DSC) loss and Cross Entropy (CE) loss are employed in every loss calculation and guide the model to predict more similar results as labels. DSC loss is region-based loss, aims to minimize the mismatch or maximize the overlap region between the label and predicted partition. CE is a distribution-based loss that is designed to minimize the distribution difference between predictions and targets.

\subsection{Implementation details}
Convolutional neural networks are data-driven and require large amounts of data to extract semantic information and construct feature representations, so we employed a variety of data augmentation strategies, including: 1) random rotation between -30 degrees and 30 degrees; 2) random vertical flipping and horizontal flipping; 3) randomly adding Gaussian noise with 50\% probability; 4) random Gaussian smoothing with 50\% probability; 5) random contrast adjustment with 50\% probability; 6) randomly cropping out a region with a shape of 16$\times$320$\times$320, where 16 represents the slice depth, and 320$\times$320 indicates the visual size; 7) Normalize the input scan by subtracting the mean value and dividing the standard deviation value. A NVIDIA Tesla A100 (80G) GPU is deployed to run the pipeline, the optimizer is Adamw with the initial learning rate as 1e-2, CosineAnnealingWarmRestarts~\cite{SGDR} is used to adjust the learning rate. Dice Similarity Coefficient (DSC)~\cite{Dice},  Intersection-over-Union (IOU), Recall, and Precision are used as segmentation evaluation metrics over the testset.  In training, we randomly crop input scans to train the pipeline, but the input is orderly cropped into patches with the shape as 16$\times$320$\times$320 to do the test in the testing phase with the sliding window size as 8$\times$160$\times$160, and the metrics are calculated after recomposing the ordered patches to the original size, and the overlapped places are averaged.

% \subsection{Quantitative analysis}
% \begin{figure*}[hbtp]
% \centering
% \includegraphics[scale=0.2]{.//img/SHOW.pdf}
% \caption{Visualization results of bleeding area segmentation overlaid to the CT slices. A) The 3D structure generate from Label, 2D, 3d-Cascade, 3D-fullres, and 3D-lowres, respectively. B) First row is 3D-Fullres, second row is 3D-Lowres, third row is 3D-Cascade-Fullres, and the last row is 2D. Red represents TP, green represents FP, and blue represents FN.Best view in color.} \label{fig:Results_show}
% \end{figure*}

\section{Results}\label{results}

\subsection{Ablation experiment}
To show the superior performance of our cascaded model, we set up the following ablation experiment settings: 1) \textbf{2D-Model}, this setting has the same structure as the first stage of our cascaded model, except this setting deployed 2D convolution and processed 2D slices with the shape of 512$\times$512 not the sequential scan. 2) \textbf{3D-Lowres}, this setting employed the same 3D convolutional structure as the first stage of our cascaded model, but it only had a single stage, and the median voxel size and spacing of the input scan are smaller than our cascaded model, the median voxel size is 28$\times$483$\times$483, and the spacing is set as 5$\times$0.518$\times$0.518. 3)\textbf{3D-Fullres}, this setting is similar to the \textbf{3D-Lowres}, but the median voxel size is 28$\times$512$\times$512, and the spacing is 5$\times$0.488$\times$0.488. 4) \textbf{3D-Cascaded}, the proposed method in which first stage is the same as \textbf{3D-Fullres}, and the second stage is the same as \textbf{3D-Lowres}. The experiment results are shown in Table~\ref{tab:ex}.

\begin{table}[htbp]
\centering
\caption{Experiments of segmentation performance}
\setlength{\tabcolsep}{2mm}{
\begin{tabular}{ccccc}
\hline \hline
\textbf{Method}    & \textbf{DSC(\%)}     & \textbf{IOU(\%)}     & \textbf{Precision(\%)} & \textbf{Recall(\%)}  \\ \hline
3D-Cascade                & \textbf{85.66}          & \textbf{76.82} & \textbf{88.46}   & 84.11          \\ \hline 
3D-Fullres         & 85.19          & 76.22         & 88.31           & 83.54         \\
2D-Model & 85.59 & 76.81          & 88.04           & 84.43          \\
3D-Lowres          & 85.63         & 76.80          & 87.83            & \textbf{84.62} \\ \hline \hline
\end{tabular}}
\label{tab:ex}
\end{table}
\subsection{Segmentation accuracy}
From the table, we can see that our method achieved the first place in three of the four metrics and obtained overall better results, which shows that compared with the 2D method, the input voxel data allows the network to better capture the contextual information between slices and model the changes in the bleeding region, thus improving the overall segmentation accuracy; and compared with 3D-Fullres and 3D-Lowres, it can be found that the large resolution of volume data can better reveal the details of the bleeding region, thus bringing performance improvement; finally, the 3D-Cascaded employs a two-level segmentation strategy to optimize the segmentation results from coarse to fine, thus achieving a better overall segmentation effect.

\begin{equation}\begin{array}{l}
\mathbb{V}_{MAE} = \frac{ {\textstyle \sum_{i}^{N}}|V_{i} - \mathcal{V}_{i}| }{N} 		
\label{for:mae}
\end{array}\end{equation}
\subsection{Diagnostic accuracy}
\begin{figure*}[hbtp]
\centering
\includegraphics[scale=0.25]{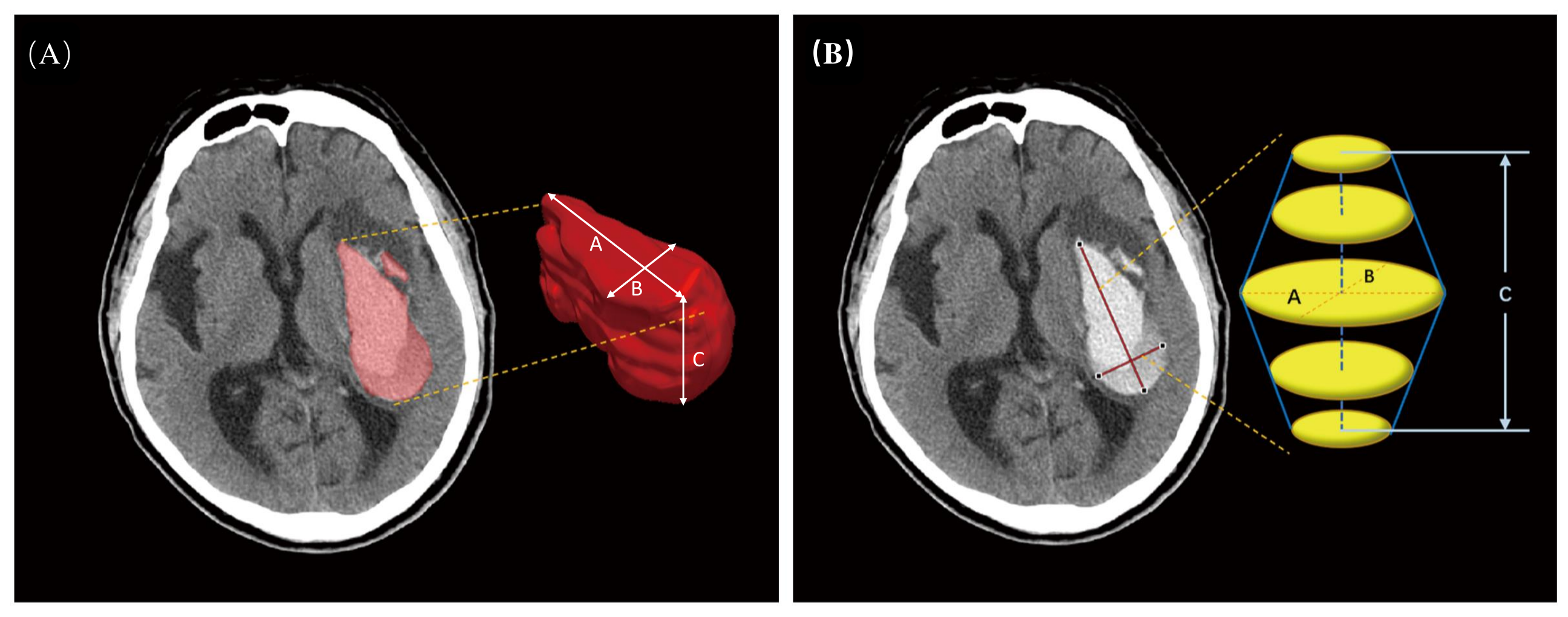}
\caption{Display of Tada formula parameters. Best view in color.} \label{fig:Tada}
\end{figure*}

Among the clinical indicators, the volume of bleeding is a very important factor. We calculate the Mean Absolute Error (MAE) of the bleeding volume between the prediction and GT to compare the diagnostic accuracy, as shown in Eq~\ref{for:mae}, where N indicates the number of samples, V and $\mathcal{V}$ represent the volume of the prediction and GT, respectively. The bleeding volume of GT are calculated by ITK-snap~\cite{itk_snap}, while the bleeding volume of prediction is by summing the non-zero pixel number of the prediction mask, and the bleeding volume of prediction is by multiplying the bleeding voxel count and the spacing of the input sample. At present, the most widely used clinical method for measuring bleeding volume is the Tada formula~\cite{ABC-1,ABC-2,ABC-3}, which was first proposed by Japanese scholars and given as
$\frac{\pi}{2\times A \times B \times C}$, where A (mm) is the maximum bleeding length, B (mm) is the maximum width perpendicular to A (mm) determined on the slice of maximal area, and C is the depth of bleeding, as shown in Fig~\ref{fig:Tada}. For the convenience of clinical practice, the Tada formula is simplified and modified into $\frac{1}{2\times A \times B \times C}$, which is widely used~\cite{2019abc}. Clinically, for IVH and IPH, physicians commonly used the Tada formula method to quickly obtain the hemorrhage volume (ml), while for SAH, due to its scattered and irregular shape, the Tada formula cannot be used to estimate the hemorrhage volume. The experimental results are shown in Table~\ref{tab:dig_results}. As can be seen, regardless of the experimental configuration, the volume difference for IVH and IPH show a great advantage over the Tada formula, with a minimum improvement of 44.66\% and a maximum improvement of 75.54\%, obtained by the proposed method. On the volume difference of SAH, the best results were achieved by the proposed method, with an improvement of 49.47\%, proving the advantages of the present method. Moreover, our method also has significant advantages in terms of the time required. A physician with 5 years of experience estimated the bleeding volume using the Tada formula in an average of 27.8 seconds per sample, while our cascaded model predict the bleeding volume in only 6.2 seconds per sample, which takes only 22.3\% of the time of Tada formula, but without losing accuracy.
\begin{table}[htbp]
\caption{Diagnostic accuracy comparision, where IPH+IVH+(IPH\&IVH) indicates represents the total of these three types of bleeding.}
\centering
\setlength{\tabcolsep}{2mm}{
\begin{tabular}{c|cc|c}
\hline
\textbf{Method} & \multicolumn{2}{c|}{\textbf{Volume (ml)}} & \textbf{Times (s)} \\ \hline
-               & IPH+IVH+(IPV\&IVH)         & SAH          & -                 \\ \hline
Tada formula    & 9.167                      & -            & 27.8              \\
2D-Model             & 5.077                      & 9.298        & 7.6               \\
3D-Lowres       & 2.487                      & 5.318        & 5.7               \\
3D-Fullres      & 2.442                      & 5.244        & 5.8               \\
3D-Cascade      & 2.242                      & 4.698        & 6.2               \\ \hline
\end{tabular}
}\label{tab:dig_results}
\end{table}

% \end{figure*}
% \begin{figure*}[hbtp]
% \centering
% \includegraphics[scale=0.35]{.//img/dig_results_time.pdf}
% \caption{Diagnostic accuracy. Best view in color.} \label{fig:dig_results}
% \end{figure*}

\section{Conculusions}\label{conculusions}
Our findings indicate that a cascaded encoder-decoder convolutional model can be trained to automatically segment intracranial bleeding in CT images and achieve promising segmentation performance when compared to human experts' hand-contouring reference. These trained cascaded encoder-decoder convolutional models may aid in clinical workflow and allow for more quantitative assessments of CT imaging modalities.

\section{Acknowledgement}\label{ACKNOWLEDGEMENT}
This research is supported by the National Key R\&D Program of China (Grant No.2018AAA0102600) and the National Natural Science Foundation of China (Grant No.62002082).

\bibliographystyle{splncs04}
\bibliography{ref.bib}
\end{document}